\documentclass[sigconf,nonacm]{acmart}
\usepackage{subcaption}
\usepackage{algorithm}
\usepackage[noend]{algorithmic}

\usepackage{textcomp}

\usepackage{makecell}

\AtBeginDocument{%
  }

\begin{document}

\title[A Dynamic-Kernel/QPacket Executable]{A Dynamic-Kernel/QPacket Executable\\ for Quantum Repeater Chains in Q2NS/ns-3}

\author{Adam Pearson}
\email{adamgerrit.pearson@unina.it}
\orcid{0009-0007-9817-0487}
\affiliation{%
  \institution{University of Naples Federico II}
  \city{Naples}
  \country{Italy}
}

\author{Marcello Caleffi}
\email{marcello.caleffi@unina.it}
\orcid{0000-0001-5726-5489}
\affiliation{%
  \institution{University of Naples Federico II}
  \city{Naples}
  \country{Italy}
}

\author{Angela Sara Cacciapuoti}
\email{angelasara.cacciapuoti@unina.it}
\orcid{0000-0002-0477-2927}
\affiliation{%
  \institution{University of Naples Federico II}
  \city{Naples}
  \country{Italy}
}

\begin{abstract}
The Quantum Internet operates on entanglement, a non-local, non-copyable, stateful network resource, which motivates protocol organization beyond classical layering. We present a first executable specialization of the Dynamic Kernel/QPacket logic from the beyond-layering protocol suite, targeting entanglement distribution over a linear quantum repeater chain. The implementation builds on \textit{Q2NS}, an ns-3-based quantum-network simulation module available through the ns-3 App Store. It realizes QPacket meta-headers with service intent and append-only action-commit stamps processed by node-local Dynamic Kernels organized as a Planner--Executor--Engine pipeline, while being deliberately scoped to an analytically verifiable service and policy. Within this scoped setting, we study node heterogeneity through a link-preparation policy that accounts for pre-distributed entanglement and uneven entanglement-generation support across nodes, including delegation via QPacket forwarding. Simulations verify analytical link-resolvability models and expose signaling load, forwarding behavior, and QPacket meta-header growth. Results show that QPacket overhead is shaped by more than just encoding, including policy choices and available network resources. Overall, this study demonstrates how the Q2NS/ns-3 substrate can support reproducible, policy-specific evaluation of quantum-native protocol-suite concepts.
\end{abstract}


\maketitle

\section{Introduction}

The Quantum Internet~\cite{CalCac-26,CacCal-26,Kim-08, DurLamHeu-17, VanSatBen-22, PirDur-19} is being designed to generate, distribute, maintain, and manipulate quantum entanglement as its key communication resource. This fundamentally departs from the classical Internet scope, where packets carry copyable information that can be buffered, duplicated, and forwarded by intermediate nodes. In
quantum networks, entanglement is non-local and stateful: local
operations at a node can affect entangled states shared with remote nodes, and
protocol decisions depend on resource descriptors such as entanglement availability, ownership, fidelity, and coherence time. As a result, quantum-network design cannot be reduced to a direct adoption of classical
Internet design principles~\cite{KozWehVan-23, CalCac-26}. 

A recently proposed response to this challenge is the \textit{beyond-layering} protocol suite~\cite{CacCal-26}, where quantum network functions are not organized into fixed vertical layers. Instead the suite dynamically composes atomic functionalities, called micro-protocols (MPs), into higher-order constructs called \textit{meta-protocols} (MePs) to realize complex network functionalities. To support this composition, a quantum packet (QPacket) carries a quantum control field, termed \textit{meta-header}, which contains the selected service intent and an append-only sequence of action-commit records, called \textit{stamps}. At each node, a local \textit{Dynamic Kernel} reads the meta-header non-destructively, enabled by an appropriate quantum encoding~\cite{CacCal-26}; it combines the packet-carried history with the node-local state; constructs a Plan of Actions (PoA), maps actions to MPs and MePs; and appends new stamps only at action-commit boundaries. In this way, service progression is certified by the packet-carried history rather than prescribed by a static layer order \cite{CacCal-26}.

However, instantiating this suite as an ns-3 executable is non-trivial. The beyond-layering suite intentionally separates the reusable architectural substrate from service-specific choices, including policies, MP libraries, routing assumptions, resource-management rules, and QPacket choices. This calls for scoped executable specializations in which these choices are made explicit, instrumented, and verified against analytical reference models. 

This paper develops such a first executable specialization using Q2NS, an ns-3-based quantum-network simulator \cite{PeaMazCal-26a} available through the ns-3 App Store~\cite{q2ns-appstore}. Q2NS
extends ns-3 with quantum-networking primitives while preserving the
ability to co-simulate quantum operations and classical communication
inside the ns-3 event-driven environment. We build on the released Q2NS substrate, without modifying its internals, and implement an executable Dynamic Kernel/QPacket specialization for end-to-end (e2e) entanglement distribution over a linear quantum repeater chain. 

The implementation separates reusable beyond-layering structures from service-specific policy logic: the abstract QPacket meta-header, append-only commit stamps, and the Planner--Executor--Engine Kernel pipeline follow the beyond-layering
design~\cite{CacCal-26}, while fixed-chain path selection, capability-dependent MP binding, and failure handling are deliberately specialized to this service and linear-chain topology, as summarized in Table~\ref{tab:scope}. The service progresses by preparing adjacent entangled links, applying entanglement swapping at intermediate repeaters, and forwarding the QPacket along the chain until delivery or failure.  Link preparation can be resolved in three ways: (i) a node can generate the required elementary entanglement locally, (ii) it can delegate generation to a neighboring node, or (iii) the required adjacent entanglement is already available as a pre-distributed resource. If none of these conditions holds, an entanglement link cannot be established under this policy and the service fulfillment fails, as detailed in Sec.~\ref{sec:nodecapspols}. 

\begin{table}[t]
\centering
\caption{Scope of the executable e2e-entanglement-distribution specialization of the beyond-layering Dynamic-Kernel.}
\label{tab:scope}
\begin{tabular}{p{0.34\linewidth}p{0.56\linewidth}}
\toprule
Concept & Implementation \\
\midrule
Service intent & e2e entanglement distribution over a quantum repeater chain \\
\cmidrule(lr){1-2}
QPacket meta-header & Service intent plus stamps \\
\cmidrule(lr){1-2}
Dynamic Kernel &
Planner--Executor--Engine pipeline \\
\cmidrule(lr){1-2}
Service-specific policy &
Capability-dependent binding to the node-local MP library\\
\cmidrule(lr){1-2}
Path selection &
Alice--Bob chain with next-neighbor forwarding only\\
\cmidrule(lr){1-2}

Physical model &
Noiseless operations; configurable propagation delay; QPacket forwarding with per-qubit emission interval \\
\bottomrule
\end{tabular}
\end{table}

The contributions of this work can be summarized as follows:
\begin{itemize}
    \item We implement an executable Dynamic Kernel and QPacket specialization in Q2NS/ns-3 for e2e-entanglement distribution over a linear quantum repeater chain, instantiating QPacket meta-headers,
    append-only commit stamps, and a Planner--Executor--Engine pipeline.
    \item We design and implement a capability-aware link-preparation
    policy for heterogeneous MP libraries, supporting local entanglement
    generation, delegation via QPacket forwarding, pre-distributed
    entanglement resources, and explicit failure outcomes.
    \item We verify the implementation against analytical link-resolvability models for deterministic and random generator placement.
    \item We instrument the simulation to measure committed actions, signaling load, QPacket forwarding behavior, runtime entanglement transmissions, and QPacket meta-header growth under different capability and resource regimes.
    \item We provide a reproducibility artifact built on Q2NS, including execution scripts, raw CSV outputs, and plotting scripts for regenerating the main figures.
\end{itemize}

\section{Dynamic Kernel Realization in Q2NS}

The executable specialization considers a linear repeater chain,
$\mathrm{Alice} - R_1 - R_2 - \cdots - \mathrm{Bob}$. In this scoped setting, link-level EPR-pair generation between
adjacent nodes, entanglement swapping, QPacket forwarding, and service termination are all exercised. Although general topologies require routing, any selected route induces a repeater chain: once the participating repeaters have been chosen, the local swapping structure is the same. 

This focus is motivated by the operational difference between classical and quantum networks: classical multi-hop forwarding cannot be directly applied in the quantum domain. In fact, an unknown quantum state cannot be copied due to the no-cloning theorem~\cite{WooZur-82}. Moreover, by the quantum measurement postulate, quantum states cannot in general be measured without disturbing them. Storage and transmission further expose qubits to decoherence effects~\cite{KozWehVan-23,IllCalMan-22}. Consequently, the classical strategy of receiving, copying, amplifying, and retransmitting information at each hop does not directly apply to quantum networks. Long-distance quantum communication is therefore organized around the distribution of entanglement resources, which applications later consume, for example through teleportation with classical correction signaling~\cite{BenBraCre-93}. In a repeater chain, adjacent EPR links are extended by entanglement swapping: a repeater performs a Bell-state measurement (BSM) on its two local halves, sends the
classical outcome onward, and thereby consumes the adjacent links to
establish a longer-range entanglement relation~\cite{BriDurCir-98,
ZukZeiHor-93}. 

This section details how the scoped specialization is realized in Q2NS. We first describe the Q2NS simulation substrate and the abstract QPacket meta-header model, then specify the Dynamic-Kernel workflow and the service-specific policies used for repeater-chain entanglement distribution. This organization exposes a reusable beyond-layering core -- the QPacket
meta-header, append-only stamps, and Planner-Executor-Engine
pipeline -- and instantiates path selection, capability-aware link-preparation policy for heterogeneous MP libraries, and failure handling as modular policy choices for repeater-chain EPR distribution.

\begin{figure*}[ht]
  \centering  
 \includegraphics[width=0.95\textwidth]{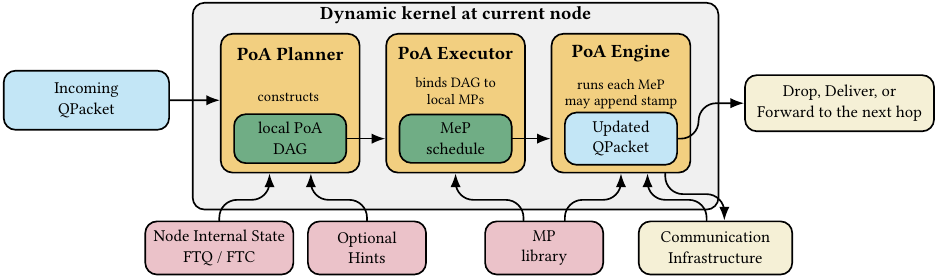}
 
  \caption{
  Dynamic-Kernel pipeline triggered by an incoming QPacket. The Planner constructs a local PoA from the QPacket meta-header and node-local state. The Executor schedules feasible actions and binds them to MPs/MePs. The Engine runs these MPs/MePs and appends stamps only at action-commit boundaries before forwarding, delivering, or dropping the QPacket. }
  \Description{
    Flow diagram of the Dynamic-Kernel processing pipeline. An incoming QPacket and node-local state enter the Planner, which constructs a local Plan of Actions. The Executor binds feasible actions to local micro-protocols (MPs) and meta-protocols (MePs), and the Engine executes them. The resulting QPacket is forwarded, delivered, or dropped, with stamps appended only when actions reach their commit boundaries.
    }
  \label{fig:dynamic-kernel-qpacket}
\end{figure*}

\subsection{Simulation Substrate}

We implement the above on top of Q2NS~\cite{PeaMazCal-26a}, an ns-3-based  quantum-network simulator available on the ns-3 App Store~\cite{q2ns-appstore}. Q2NS extends ns-3 with quantum networking primitives, while preserving ns-3's event-driven execution model and classical network stack. In particular, Q2NS introduces the \texttt{QNode} class derived from \texttt{ns3::Node} which can engage with all the classical networking simulations expected in ns-3, while also allowing transmission and reception of qubits and node-local quantum processing such as computations and measurements. There are similarly derived classes such as \texttt{QChannel} and \texttt{QNetDevice}. However, quantum networking simulations require management of quantum states and their potentially non-local nature, such as with entanglement. To address this, Q2NS introduces the \texttt{NetController}, which acts as a logically centralized controller and registry for quantum states and quantum networking objects such as \texttt{Qubits}, \texttt{QNodes}, and \texttt{QChannels}.

For more details on the architecture and performance of Q2NS, please refer to~\cite{PeaMazCal-26a} and the Q2NS repository~\cite{PeaMazCal-26b}. In this work, we build entirely on the released Q2NS functionalities, without modifying its simulation core. The Dynamic-Kernel/QPacket logic is implemented as an additional specialization layer on top of the Q2NS substrate, through new C++ data structures, service-policy logic, and instrumentation specific to the executable specialization studied here.

\subsection{Dynamic Kernel and QPackets}

The beyond-layering suite is centered on two abstractions: the QPacket and the Dynamic Kernel~\cite{CacCal-26}. QPackets carry a payload of entangled qubits and a quantum control field, the meta-header, carrying the service intent and committed-action history. Upon receiving an incoming QPacket, the node runs its \textit{Dynamic Kernel}, which advances the service intent by using the node internal state and the in-band meta-header carried by the QPacket. The Kernel acts as the node-local orchestration and execution environment: it plans candidate actions, composes the corresponding MPs into MePs, invokes the resulting realizations, and
records completed action outcomes as append-only stamps at action-commit boundaries, as illustrated in Fig.~\ref{fig:dynamic-kernel-qpacket}.
\begin{enumerate}
    \item \textit{PoA Planner}: The Planner constructs a local and speculative Plan of Actions (PoA) from the service intent, the accumulated stamps, and the node internal state. The internal state includes FTC and FTQ, respectively capturing classical reachability and the node entanglement inventory. Optional control-plane hints may refine the local planning policy. The PoA is represented as a directed acyclic graph (DAG) of candidate actions and dependency constraints.
    \item \textit{PoA Executor}: The Executor binds candidate actions to concrete MP/MeP realizations. An MP implements an atomic network functionality, while an MeP is a higher-order construct obtained by dynamically composing MPs to realize a more complex functionality. Using the node's MP library, local dependencies, and node capability constraints, the Executor selects feasible actions and produces a dependency-respecting runnable schedule.
    \item \textit{PoA Engine}: The Engine executes the scheduled MP/MeP realizations and certifies service progress only at action-commit boundaries via stamps. Each committed action appends the corresponding stamp to the QPacket meta-header. Local execution terminates by forwarding the packet to the next hop, delivering the service, or dropping the packet. After a forward, the next node runs its own Dynamic Kernel.
\end{enumerate}

The QPacket is realized as a C++ struct containing the service intent and accumulated stamps. Thus, the QPacket meta-header is modeled at a logical level, rather than as a concrete quantum-state encoding. To model QPacket forwarding, we assign the QPacket a logical qubit-equivalent size, denoted as $n_{\text{qpacket}}$. At each forwarding event, $n_{\text{qpacket}}$ is  computed by equating the number of qubits with the number of classical bits needed to represent the current variables of the QPacket structure. Thereby the model charges one qubit-equivalent transmission per represented bit. By assuming serial emission over the quantum channel with qubit emission interval of $t_{\text{emit}}$ and a propagation delay of $t_{\text{delay}}$\footnote{Single-qubit transmissions, used in $\textbf{MP}_{\textbf{GEN}}$, propagate on Q2NS \texttt{QChannels} with propagation delay $t_{\text{delay}}$.}, the total QPacket forwarding time is $t_{\text{delay}} + (n_{\text{qpacket}} - 1)t_{\text{emit}}$. We chose $t_{\text{delay}} = 0.5$ ms, corresponding to a long, but experimentally demonstrated~\cite{wengerowsky2019entanglement} 100 km of fiber, with a 5~\textmu s/km propagation delay. We also chose $t_{\text{emit}} = 1$ ns, corresponding to an effective 1 GHz photon-emission rate, consistent with high-rate
single-photon sources~\cite{lounis2005single,senellart2017high}. These are used as
representative timing parameters, however the focus of this study is on scaling trends that are largely independent of these choices. Overall, this model serves as a logical baseline and not as  a physical QPacket encoding
proposal. Concrete implementations may introduce additional overheads from
physical encoding choices, synchronization, hardware constraints, or
error correction. Thus, the measured forwarding delays quantify meta-header growth trends under the abstract model, not absolute physical transmission costs. Classical messages are sent as UDP/IPv4 packets under 30 bytes over ns-3 point-to-point links also with propagation delay $t_{\text{delay}} = 0.5$ ms and a 1 Gbps data rate.

Dynamic-Kernel processing and individual MP execution are instantaneous in simulation time. This isolates QPacket forwarding and meta-header growth effects. Future work will incorporate realistic processing times, particularly important in noisy scenarios where the quality of distributed entanglement decreases over time.

Algorithm~\ref{alg:qkernel-workflow} summarizes the Dynamic-Kernel workflow for the executable repeater-chain specialization. Each repeater prepares missing adjacent links, swaps between its left and right neighbors, and forwards the updated QPacket. Table~\ref{tab:action-mp-communication} details the mapping from actions to MPs, simulated exchanges,
and state/QPacket updates.

We record Dynamic-Kernel metrics through custom counters maintained by a
\texttt{DynamicKernelManager} object and by per-node runtime state. These metrics include committed actions, delegated-generation requests, signaling messages, QPacket forwarding events, failure depth, and QPacket meta-header growth. They are implemented as custom counters, rather than ns-3 TraceSources, because they are internal to the service-specific Dynamic-Kernel logic instantiated in this work. A future Q2NS module could expose the stable subset of these events as ns-3 TraceSources.

\begin{algorithm}[!htp]
\caption{Dynamic Kernel workflow for the executable repeater-chain specialization}
\label{alg:qkernel-workflow}

{\fontsize{8.5pt}{9pt}\selectfont
\begin{algorithmic}[1]

\STATE Configure network, MP libraries, and pre-distribute entanglement
\STATE Inject QPacket at Alice with intent
$\mathrm{EPR\_DISTRIBUTE}(\mathrm{Alice},\mathrm{Bob})$

\STATE
\STATE \textbf{ReceivePacket}$(\mathrm{QPacket})$
\COMMENT{Trigger Dynamic Kernel}
\STATE $\mathrm{PoA}
       \gets
       \textsc{Planner}(\mathrm{QPacket},\mathrm{FTQ},\mathrm{FTC})$
\STATE $\mathrm{ExecPlan}
       \gets
       \textsc{Executor}(\mathrm{PoA},\mathrm{MPs})$
\STATE $\mathrm{QPacket}_{\mathrm{updated}}
       \gets
       \textsc{Engine}(\mathrm{ExecPlan},\mathrm{MPs},\mathrm{QPacket})$

\STATE
\STATE \textbf{Planner}$(\mathrm{QPacket},\mathrm{FTQ},\mathrm{FTC})$
\COMMENT{Build PoA}
\STATE $A \gets$ first participant in QPacket intent; $B \gets$ last participant
\STATE $\mathrm{left} \gets$ left neighbor; $\mathrm{right} \gets$ right neighbor

\IF{\textsc{LINK\_PREP}$(\mathrm{left})$ failed in
    $\mathrm{QPacket.stamps}$}
    \STATE Add \textsc{LINK\_PREP}$(\mathrm{left})$
           with delegation forbidden
    \STATE Add
    \textsc{ACT\_HOLD}(
      \textsc{SWAP}$(\mathrm{prev(left)},\mathrm{this\ node})$;
      \textsc{certifying})
\ENDIF

\IF{$\mathrm{this\ node}=B$}
    \STATE Add \textsc{CORRECT\_AND\_ACK}$(A)$
    \STATE Add \textsc{ACT\_DELIVER}

\ELSIF{$\mathrm{this\ node}=A$}
    \IF{$\mathrm{right}\notin\mathrm{FTQ}$}
        \STATE Add \textsc{LINK\_PREP}$(\mathrm{right})$
    \ENDIF
    \STATE Add \textsc{ACT\_FORWARD}$(\mathrm{right})$
    \STATE Add
    \textsc{ACT\_HOLD}(\texttt{CompletionAck};
                       \textsc{release-only})

\ELSE
    \IF{$\mathrm{right}\notin\mathrm{FTQ}$}
        \STATE Add \textsc{LINK\_PREP}$(\mathrm{right})$
    \ENDIF
    \STATE Add \textsc{SWAP}$(\mathrm{left},\mathrm{right})$
    \STATE Add \textsc{ACT\_FORWARD}$(\mathrm{right})$
\ENDIF

\STATE Add action-level dependencies
\STATE \textbf{return} $\mathrm{PoA}$

\STATE
\STATE \textbf{Executor}$(\mathrm{PoA},\mathrm{MPs})$
\COMMENT{Build ExecPlan}
\STATE $\mathrm{canGen}
       \gets
       (\mathbf{MP}_{\mathbf{GEN}}\in\mathrm{MPs})$

\IF{\textbf{not} $\mathrm{canGen}$
    $\land$
    (\textsc{LINK\_PREP} with delegation forbidden)}
    \STATE \textbf{return} $[\,]$
\ENDIF

\FOR{each action in $\mathrm{PoA}$}
    \IF{action is \textsc{ACT\_HOLD}}
        \STATE Add as \textsc{service-soft-action}
        \COMMENT{Soft state}

    \ELSIF{action is \textsc{LINK\_PREP}$(v)$
           and \textbf{not} $\mathrm{canGen}$}
        \STATE Add as \textsc{non-bindable}
        \COMMENT{Engine records a failure stamp}
        \STATE Attach \textsc{runtime-soft-hold}
               until $v\in\mathrm{FTQ}$

    \ELSE
        \STATE Bind action according to
               Table~\ref{tab:action-mp-communication}
        \STATE Add bound action to $\mathrm{ExecPlan}$
    \ENDIF
\ENDFOR

\STATE \textbf{return} $\mathrm{ExecPlan}$

\STATE
\STATE \textbf{Engine}$(\mathrm{ExecPlan},\mathrm{MPs},\mathrm{QPacket})$
\COMMENT{Run ExecPlan}

\IF{$\mathrm{ExecPlan}=[\,]$}
    \STATE Commit \textsc{ACT\_DROP} and terminate the request
\ENDIF

\FOR{each scheduled item in $\mathrm{ExecPlan}$}
    \IF{item is \textsc{service-soft-action}}
        \STATE Install local service-level soft state
        \IF{mode is \textsc{certifying}}
            \STATE Append stamp after release
        \ENDIF

    \ELSIF{item is \textsc{non-bindable}}
        \STATE Append failure stamp for the action
        \COMMENT{No MP is invoked}

    \ELSIF{item has an unsatisfied
           \textsc{runtime-soft-hold}}
        \STATE Install local runtime-level soft state
        \COMMENT{No stamp}

    \ELSE
        \STATE Execute the bound MPs according to
               Table~\ref{tab:action-mp-communication}
        \STATE Apply local state-transition effects
        \STATE Append stamp after the action reaches its commit boundary
    \ENDIF
\ENDFOR

\STATE \textbf{return} $\mathrm{QPacket}$

\end{algorithmic}
}
\end{algorithm}

\begin{table*}[t]
\centering
\caption{
Executor bindings and Engine behavior for each action in the executable repeater chain specialization
}
\label{tab:action-mp-communication}
\footnotesize
\setlength{\tabcolsep}{4pt}
\newcommand{\cellwrap}[1]{%
  \begin{minipage}[t]{\linewidth}%
  \raggedright
  #1%
  \end{minipage}%
}
\begin{tabular}{@{}
p{0.17\textwidth}
p{0.09\textwidth}
p{0.68\textwidth}
@{}}
\toprule
\textbf{Planner Action} &
\textbf{Executor MeP} &
\textbf{Engine Implementation} \\ 
\midrule
\textsc{LINK\_PREP}$(x)$
&
$\textbf{MP}_{\textbf{SYN}}$
&
\cellwrap{
1. Send link-preparation packet for generation from this node to $x$\\
2. Receive acknowledgment from $x$
}\\
\cmidrule(lr){2-3}
&
$\textbf{MP}_{\textbf{GEN}}$
&
\cellwrap{
1. Generate EPR pair\\
2. Send half of EPR pair to $x$\\
3. Apply \textsc{LINK\_PREP} state effect: add $x$ to this node's FTQ and this node to $x$'s FTQ\\
}
\\
\cmidrule(lr){2-3}
\multicolumn{3}{@{}p{\dimexpr\textwidth-2\tabcolsep\relax}@{}}{%
\textbf{Note:}
If node lacks $\textbf{MP}_{\textbf{GEN}}$, then \textsc{non-bindable}. The Engine appends a failure stamp for \textsc{LINK\_PREP}$(x)$ and installs a runtime soft hold waiting for reciprocal link preparation.
}\\
\addlinespace[1pt]
\midrule
\textsc{SWAP}$(x,y)$
&
\makecell[tl]{$\textbf{MP}_{\textbf{QP}}$}
&
\cellwrap{
1. Wait until the runtime guard is satisfied: both adjacent FTQ entries are available.\\
2. BSM on two locally stored ebit halves (one from $x$ and one from $y$)\\
3. Remove $x$ and $y$ from $\text{FTQ}$\\
4. Aggregate BSM outcome with classical correction packet received by $x$
}\\
\cmidrule(lr){2-3}
&
\makecell[tl]{$\textbf{MP}_{\textbf{SIG}}$}
&
\cellwrap{
Send aggregated classical correction packet to $y$, triggering $y$ to change $\text{FTQ}$ entry from this node to $x$
}
\\
\addlinespace[1pt]
\midrule
\textsc{CORRECT\_AND\_ACK}$(x)$
&
\makecell[tl]{$\textbf{MP}_{\textbf{QP}}$}
&
\cellwrap{
Apply corrections when aggregated correction packet arrives
}\\
\cmidrule(lr){2-3}
&
\makecell[tl]{$\textbf{MP}_{\textbf{SIG}}$}
&
\cellwrap{
Send \texttt{CompletionAck} packet to $x$, releasing \textsc{ACT\_HOLD}(\texttt{CompletionAck}) at $x$ and applying the corresponding update of $\text{FTQ}_{x}$
}\\
\addlinespace[1pt]
\midrule
\textsc{ACT\_FORWARD}$(x)$
&
$\textbf{MP}_{\textbf{FW}}$
&
\cellwrap{
Forward QPacket to $x$, triggering Dynamic Kernel pipeline at $x$
}
\\
\addlinespace[1pt]
\midrule
\textsc{ACT\_HOLD}(trigger; mode)
&
None
&
\cellwrap{
Install service-level soft state, released by trigger. A stamp is only appended if mode is certifying. No MP invoked
}\\
\addlinespace[1pt]
\midrule
\textsc{ACT\_DELIVER}
&
None
&
\cellwrap{
Service terminates successfully, no further QPacket forwarding\\
}
\\
\addlinespace[1pt]
\midrule
\textsc{ACT\_DROP}
&
None
&
\cellwrap{
Drop QPacket. Service terminates in failure
}
\\
\bottomrule
\end{tabular}
\end{table*}

\subsection{Node Capabilities and Delegation Policy}
\label{sec:nodecapspols}

Realistic quantum networks are expected to contain nodes with heterogeneous capabilities. In the beyond-layering suite \cite{CacCal-26}, this heterogeneity is naturally represented through node-local MP libraries: each node can bind and invoke only the MPs exposed by its own library. In the present repeater-chain specialization, we instantiate
this heterogeneity through entanglement-generation capability. Some nodes expose $\textbf{MP}_{\textbf{GEN}}$, while others do not. In the latter case, a \textsc{LINK\_PREP} action cannot be
bound locally.

This setting stresses an important feature of the beyond-layering suite: the absence of a local MP does not by itself invalidate the service progression. 
Indeed, the Dynamic Kernel pipeline naturally adapts to missing local
capabilities while preserving local autonomy and the single-writer discipline of the QPacket meta-header. If the current QPacket holder cannot generate entanglement locally, the Executor marks
\textsc{LINK\_PREP} as non-bindable. The Engine then invokes no MP and
records the failed local realization by appending a failure stamp for
\textsc{LINK\_PREP}. The QPacket is then forwarded according to the local PoA. The receiving node runs its own Dynamic Kernel, reads the updated stamp history, and may schedule the reciprocal \textsc{LINK\_PREP} using its own local MP library. For a requested adjacent link between the current QPacket holder $x$ and its neighbor $y$, the implemented policy is as follows:
\begin{enumerate}
    \item If $x$'s FTQ already contains an entanglement resource for $y$,
    no \textsc{LINK\_PREP} action is added to the PoA.
    \item Otherwise, the Planner adds \textsc{LINK\_PREP}$(y)$ to the
    local PoA. 
    \begin{itemize}
    \item If $x$ exposes $\textbf{MP}_{\textbf{GEN}}$, the Executor binds \textsc{LINK\_PREP}$(y)$ according to
    Table~\ref{tab:action-mp-communication}. The resulting MeP synchronizes the endpoints and generates an EPR pair at $x$, which sends one half to $y$. Upon completion, the Engine applies
    the corresponding FTQ updates and appends a committed
    \textsc{LINK\_PREP} stamp to the QPacket meta-header.
    \item If $x$ lacks $\textbf{MP}_{\textbf{GEN}}$, the Executor marks \textsc{LINK\_PREP} as non-bindable, by scheduling it with an empty MeP. This signals to the Engine to invoke no MP and append a failure stamp. The QPacket is then forwarded to $y$ when \textsc{ACT\_FORWARD}$(y)$ is
    present in the local PoA. Subsequently, the Engine at $x$ installs a local runtime soft hold waiting for reciprocal link preparation. This hold is a local state under a release-only mode, meaning it does not append a stamp.
    \item When $y$ receives the forwarded QPacket, it runs a new Dynamic-Kernel pipeline. From the failure stamp, the Planner knows to add the reciprocal \textsc{LINK\_PREP}$(x)$, with delegation forbidden. If $y$ also lacks $\textbf{MP}_{\textbf{GEN}}$, the service terminates with \textsc{ACT\_DROP}. If $y$ exposes $\textbf{MP}_{\textbf{GEN}}$, it realizes the link locally and sends one EPR-pair half to $x$, releasing $x$'s runtime soft hold. After \textsc{LINK\_PREP}, $y$ installs a local service soft hold waiting for signaling from $x$ that it has completed its SWAP and, according to the certifying mode, stamps this locally at $y$ since $x$ no longer has the QPacket. 
\end{itemize}
    \item If the current QPacket holder cannot progress as no feasible local continuation remains, the service terminates in failure with \textsc{ACT\_DROP}.
\end{enumerate}

Our implementation is aligned with the invariants of the beyond-layering suite. Beyond the aforementioned \textit{local autonomy} and \textit{single-writer discipline}, it also preserves \textit{in-band coordination}, because the service context is transferred through the QPacket meta-header and its append-only stamp
history, rather than through global synchronization or an out-of-band speculative plan. And, it preserves \textit{dynamic composition}, since each receiving node recomputes its continuation locally from the stamp history and composes its available MPs into an appropriate MeP.

This specialization also distinguishes two architectural classes of
local soft state, which differ in origin and semantics. The first class
consists of explicit service-level soft actions, represented by
\textsc{ACT\_HOLD}. These actions are inserted by the Planner when the
waiting condition is part of the service semantics derived from the
QPacket meta-header. This information must be encoded in the PoA, since the Executor receives the PoA and the local MP library but does not directly inspect the QPacket meta-header. We type \textsc{ACT\_HOLD} according to its release behavior. A
\textit{release-only} hold is released by its trigger and does not append a
stamp. This is the case, for example, of
\textsc{ACT\_HOLD}$(\texttt{CompletionAck};\textsc{release-only})$, which lets the service initiator wait for the end-to-end completion acknowledgment. A \textit{certifying} hold is also installed as local soft state, but its trigger carries evidence that another node has completed an action whose stamp must be written by the current holder of the
authoritative QPacket meta-header. After validating the trigger and the associated consistency conditions, the Engine releases the hold and appends the corresponding stamp. Thus,
\textsc{ACT\_HOLD}$(\texttt{SwapReport};\textsc{certifying})$ can be used when a neighboring node performs a local \textsc{SWAP} after the QPacket has already been forwarded, while the current QPacket holder remains responsible for appending the \textsc{SWAP} stamp.

The second class consists of runtime soft holds. These are not Planner
actions. They are introduced by the Executor when a scheduled realization cannot proceed until an execution-time condition becomes true, such as reciprocal link preparation after a failed local
\textsc{LINK\_PREP}, or FTQ availability before \textsc{SWAP}. Thus, \textsc{ACT\_HOLD} communicates a service-level waiting requirement from the Planner to the Executor, whereas runtime soft holds are local execution guards attached to scheduled items. Runtime soft holds are
always local non-stamping state.

Soft-state release is trigger-based. In the simulator, incoming events are typed so that QPacket arrivals are distinguished from soft-state triggers and side information. Only a QPacket starts a new Dynamic-Kernel pipeline. By contrast, a soft-state trigger is a classical or quantum event, such as a \texttt{CompletionAck}, a reciprocal link-preparation event, a received EPR-pair half, or a correction packet. Such a trigger is delivered through the signaling path and matched against locally
installed soft states. It does not start a new authoritative Kernel pipeline. A trigger notifies the Engine that a local runtime condition may have changed. The Engine then applies the local state-transition effect according to the action/MeP semantics and re-evaluates the guard associated with the hold. 

\begin{table}[t]
\centering
\caption{Generator-capability placements used in the case study. $G_i$ is whether node $i$ has $\textbf{MP}_{\textbf{GEN}}$, indexed $0,\ldots,N-1$, and $n_{\mathrm{gen\text{-}unres}}$ counts links where both endpoints lack $\textbf{MP}_{\textbf{GEN}}$.}
\label{tab:capability-gap-counts}
\footnotesize
\setlength{\tabcolsep}{4.5pt}
\begin{tabular}{@{}
p{0.21\linewidth}
p{0.2\linewidth}
p{0.5\linewidth}
@{}}
\toprule
\textbf{Placement} &
\textbf{$G_i$} &
\textbf{$n_{\mathrm{gen\text{-}unres}}$} \\
\midrule
\texttt{all}              & 1 $\forall i$                   & $0$ \\
\texttt{every2}           & $i \equiv 0 \pmod{2}$           & $0$ \\
\texttt{every3}           & $i \equiv 0 \pmod{3}$           & $\lfloor \frac{N}{3} \rfloor$ \\
\texttt{every4}           & $i \equiv 0 \pmod{4}$           & $2\lfloor \frac{N-1}{4} \rfloor +\max\!\left(0,((N-1)\bmod 4)-1\right)$ \\
\texttt{middleOnly}       & $i=\lfloor N/2\rfloor$          & $\max(0,N-3)$ \\
\texttt{finalBottleneck}  & $i+2<N$                         & $1$ \\
\texttt{random}           & Bernoulli($g$)                  & Random variable $n_{\mathrm{gen\text{-}unres}}(G)$ \\
\bottomrule
\end{tabular}
\end{table}

Under this policy, a physical link is resolvable when it is already
present in FTQ or when at least one endpoint exposes $\textbf{MP}_{\textbf{GEN}}$.
This gives a link-resolvability condition that admits a simple analytical success probability characterization, which we compare directly with simulation outcomes in Sec.~\ref{sec:res} for several generator-capability configurations.

Let $G_i=1$ indicate that node $i\in\{0,\ldots,N-1\}$ has $\textbf{MP}_{\textbf{GEN}}$ and $n_{\mathrm{gen\text{-}unres}}(G)$ be the number of generator-unresolvable physical links, i.e. those such that $G_i=0 \ \land\ G_{i+1}=0$. These are the physical links that cannot be prepared by either endpoint during service fulfillment. For deterministic capability placements, $G$ and $n_{\mathrm{gen\text{-}unres}}(G)$ are fixed by the placement. For random capability placement, each $G_i$ is independently sampled with $\Pr(G_i=1)=g$. The specific placements we investigate in this work and their $n_{\mathrm{gen\text{-}unres}}(G)$ values are presented in Table~\ref{tab:capability-gap-counts}.

When there is no pre-distributed entanglement, i.e., $p=0$, the service is only successful if $n_{\mathrm{gen\text{-}unres}}(G)=0$. For random capability placement, this is the probability that no two non-generator nodes are adjacent
\[
P_{\mathrm{succ}}(N,g,0)
=
\sum_{k=0}^{\lfloor (N+1)/2 \rfloor}
\binom{N-k+1}{k}
(1-g)^k g^{N-k}.
\]

If each link is independently pre-distributed an entangled pair with probability $p > 0$, then success relies on all $n_{\mathrm{gen\text{-}unres}}(G)$ generator-unresolvable links being pre-distributed, resulting in
\[
P_{\mathrm{succ}}(G,p)=p^{n_{\mathrm{gen\text{-}unres}}(G)}.
\]

For random capability placement, the success probability averages the fixed-placement expression over generator placements.

\section{Results}
\label{sec:res}

\begin{figure}[t]
  \centering
  \includegraphics[width=0.95\columnwidth]{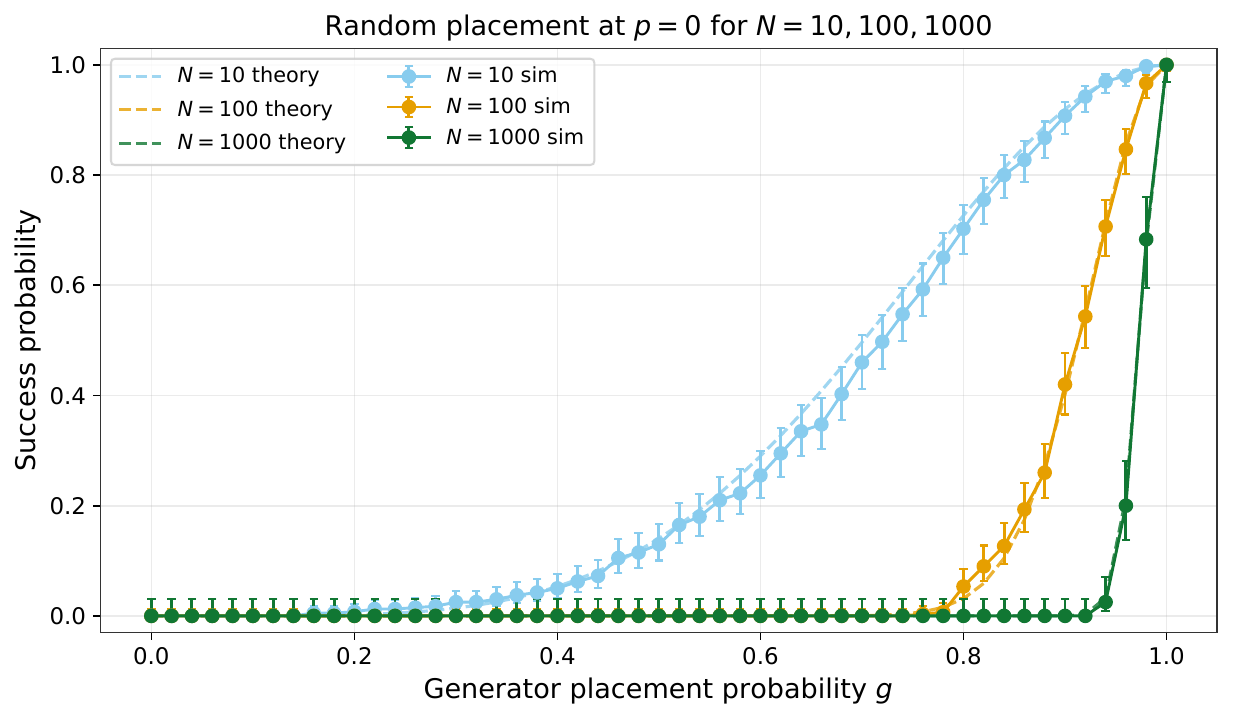}
  \caption{
  Success probability of random generator placement with probability $g$ without pre-distributed entanglement. Markers show Q2NS simulation results with 95\% binomial confidence intervals; dashed curves show the analytical probability that no two adjacent nodes both lack $\textbf{MP}_{\textbf{GEN}}$.
  }
  \Description{
    Line plot of success probability against generator-placement probability for chains of 10, 100, and 1000 nodes with no pre-distributed entanglement. Analytical curves and simulation markers agree. The transition from failure to success shifts closer to a generator probability of one as the chain becomes longer.
    }
  \label{fig:random-p0}
\end{figure}

\begin{figure}[t]
  \centering
  \includegraphics[width=0.95\columnwidth]{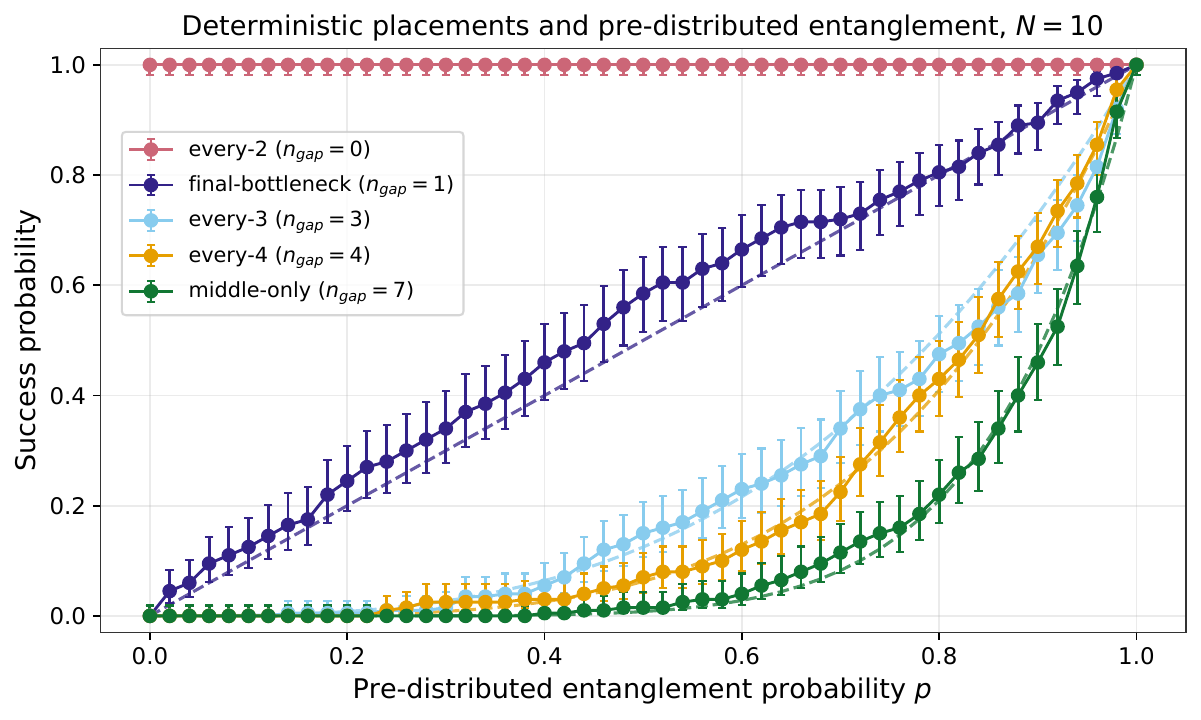}
  \caption{
  Success probability for deterministic generator-capability placements as a function of the pre-distributed entanglement probability $p$ for $N=10$. Markers show Q2NS simulation results with 95\% binomial confidence intervals, while dashed curves show the analytical probability $p^{n_{\mathrm{gen\text{-}unres}}}$.
  }
\Description{
    Line plot of success probability against pre-distributed entanglement probability for five deterministic generator-capability placements in a 10-node chain. Simulation markers agree with the analytical curves. Placements with more generator-unresolvable links require substantially greater pre-distribution probability.
    }
  \label{fig:deterministic-predistribute}
\end{figure}

This section evaluates the executable specialization along three axes. First, we verify the behavior of the implementation by comparing simulated outcomes with the analytic link-resolvability models developed in Sec.~\ref{sec:nodecapspols}. Second, we examine how signaling load and success vary across capability placements and pre-distributed entanglement. Third, we quantify QPacket meta-header growth and the accumulated QPacket forwarding delay.

Data was collected for a range of network sizes, configurations of node capabilities that stress particular failure modes, and amounts of pre-distributed entanglement. Simulations were run on a MacBook with an Apple M5 chip and 16 GB Unified Memory. The results presented here verify the specialized Dynamic-Kernel/QPacket implementation, illustrate the types of analyses supported by Q2NS/ns-3, and identify key directions for future modularization and extensions. All data is available in this work's artifact~\cite{Q2NSBeyondLayeringArtifact-26}.

The main observations of these simulations are:
\begin{enumerate}
    \item The executable outcomes match the analytical success probabilities predicted by the link-resolvability model for the considered capability placements and pre-distributed entanglement levels.
    \item Under the implemented policy and linear-chain topology, success appears more sensitive to generator probability $g$ than to pre-distributed entanglement probability $p$.
    \item Append-only QPacket stamps induce linear terminal meta-header
    growth with chain length, while repeated forwarding of the growing meta-header produces a quadratic accumulated forwarding delay.
\end{enumerate}

\subsection{Heterogeneous Node Capability Operation without Pre-distributed Entanglement}
\label{sec.3.1}

We consider heterogeneous node capabilities, modeled by node-dependent
MP libraries, in the worst-case autonomous regime. Specifically, we set the pre-distributed entanglement probability to $p=0$, so that no adjacent entanglement resource is available before the request, and we do not use control-plane hints for capability-aware placement or resource
activation. Unlike the general beyond-layering architecture in~\cite{CacCal-26}, where optional hints may assist local decisions without changing the suite semantics or its correctness, the Dynamic Kernel here operates fully autonomously only from the QPacket meta-header and node-local state. This choice deliberately tests the hint-free regime: the observed agreement with the analytical model confirms that hints are not required for
correct execution, but only for possible policy optimization. 

As a simple first verification, the simulated configurations matched the theoretical analysis in Sec.~\ref{sec:nodecapspols}: in the absence of pre-distributed entanglement, a service request succeeds
only if every physical link in the chain has at least one endpoint exposing $\textbf{MP}_{\textbf{GEN}}$. Accordingly, \texttt{all} and \texttt{every2} always succeed, while all other placements always fail. This is evident from the rightmost point of Fig.~\ref{fig:random-p0}, representing 100\% success when \texttt{all} nodes have $\textbf{MP}_{\textbf{GEN}}$, and the leftmost point of Fig.~\ref{fig:deterministic-predistribute}, representing either 100\% or 0\% success at $p=0$.

For random capability placement, Fig.~\ref{fig:random-p0} matches the closed-form success probability for a range of network sizes. The figure also shows that, as the chain length $N$ increases, success requires larger generator probability $g$. This is expected: for fixed $g$, increasing $N$ increases the probability that at least one generator-unresolvable physical link appears in the chain. The random model should therefore be interpreted as a stress model for autonomous operation. If generator-capable nodes could be placed intentionally in a known linear chain, the failure modes observed here can be avoided by construction. However, the random model can also be interpreted as capturing uncertain or time-varying generation availability, where $g$ is the probability that a node can execute $\textbf{MP}_{\textbf{GEN}}$ when the request arrives. Such availability may vary because of resource contention, source dead time, memory occupancy,  local policy constraints, etc.

Overall, Fig.~\ref{fig:random-p0} quantifies how quickly autonomous
link-resolvability (worst-case scenario) degrades with network size when neither pre-distributed entanglement nor control-plane hints are available. Control-plane hints, planned generator placement, waiting policies for temporarily unavailable generators, or alternative service policies would change the cost and timing model. Exploring these richer policy spaces is a natural next step enabled by the Dynamic-Kernel/QPacket executable developed in this work.

\begin{figure}[t]
  \centering
  \includegraphics[width=0.95\columnwidth]{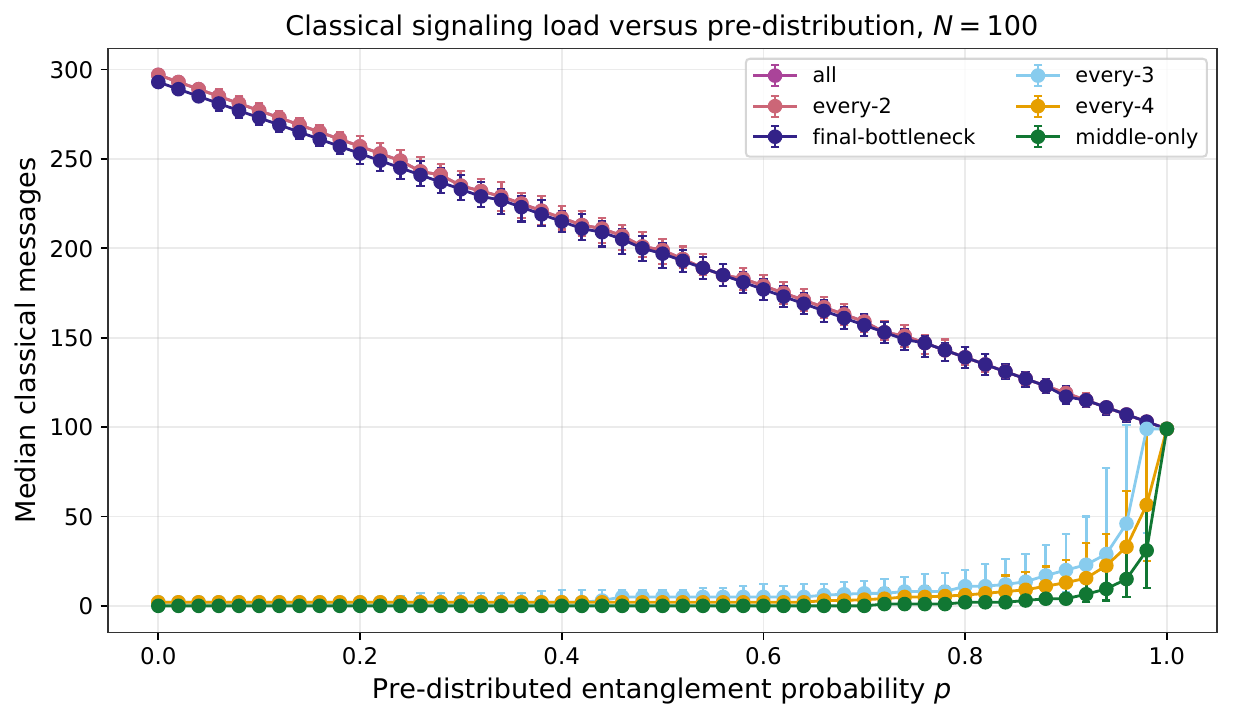}
  \caption{
  Classical signaling load versus pre-distributed entanglement probability for $N=100$. Increasing $p$ reduces \textsc{LINK\_PREP} associated signaling, but can enable further progression before failure for unresolvable link placements.
  }
    \Description{
    Line plot of median classical-message count against pre-distributed entanglement probability for six generator-capability placements in a 100-node chain. Successful placements exchange fewer messages as pre-distribution increases, while initially failing placements may exchange more messages when additional resources allow the request to progress farther before termination.
    }
  \label{fig:comm-load-predistributed}
\end{figure}

\begin{figure}[t]
  \centering
  \includegraphics[width=0.9\columnwidth]{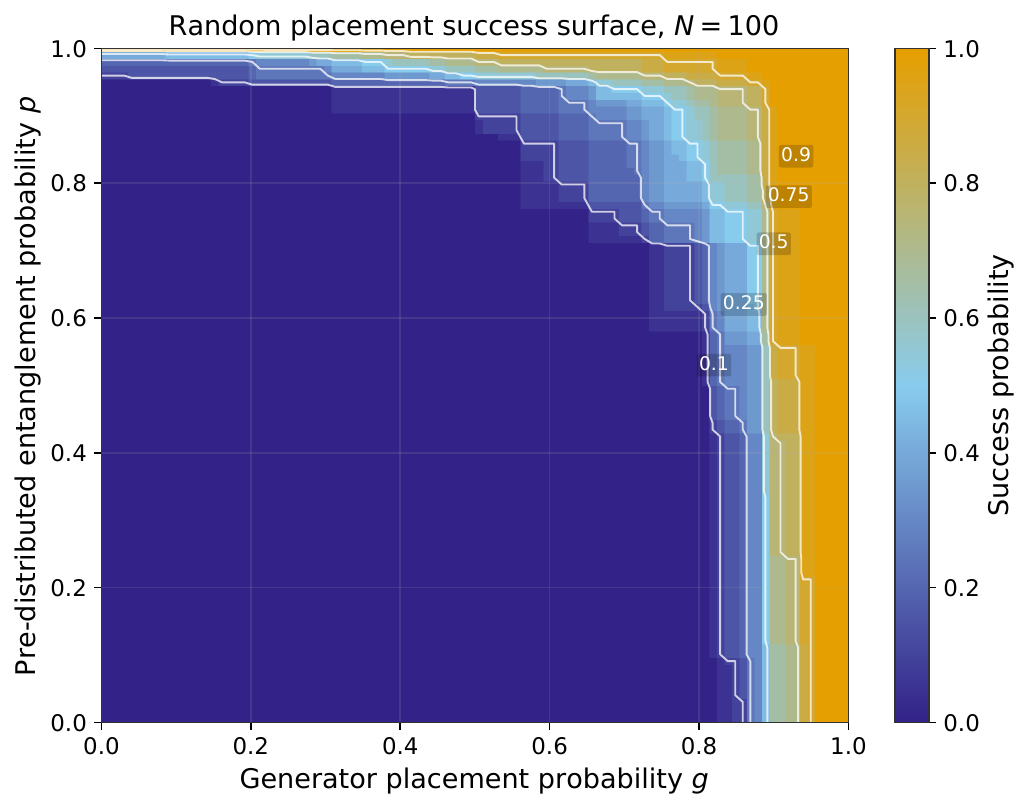}
  \caption{
  Success probability under random generator placement with probability $g$ and pre-distributed entanglement with probability $p$ for $N=100$. For each pair $(g,p)$, the color indicates the empirical success probability. Contours interpolate samples at iso-success levels $0.1$, $0.25$, $0.5$, $0.75$, and $0.9$. 
  }
  \Description{
    Heatmap of success probability for a 100-node chain as a function of generator-placement probability and pre-distributed entanglement probability. Success remains low across most of the parameter space and rises sharply at high generator probability. Contour lines mark success probabilities of 0.1, 0.25, 0.5, 0.75, and 0.9.
    }
  \label{fig:random-heatmap}
\end{figure}

\begin{figure*}[t]
  \centering

  \begin{subfigure}[t]{0.33\textwidth}
    \centering
    \includegraphics[width=\linewidth]{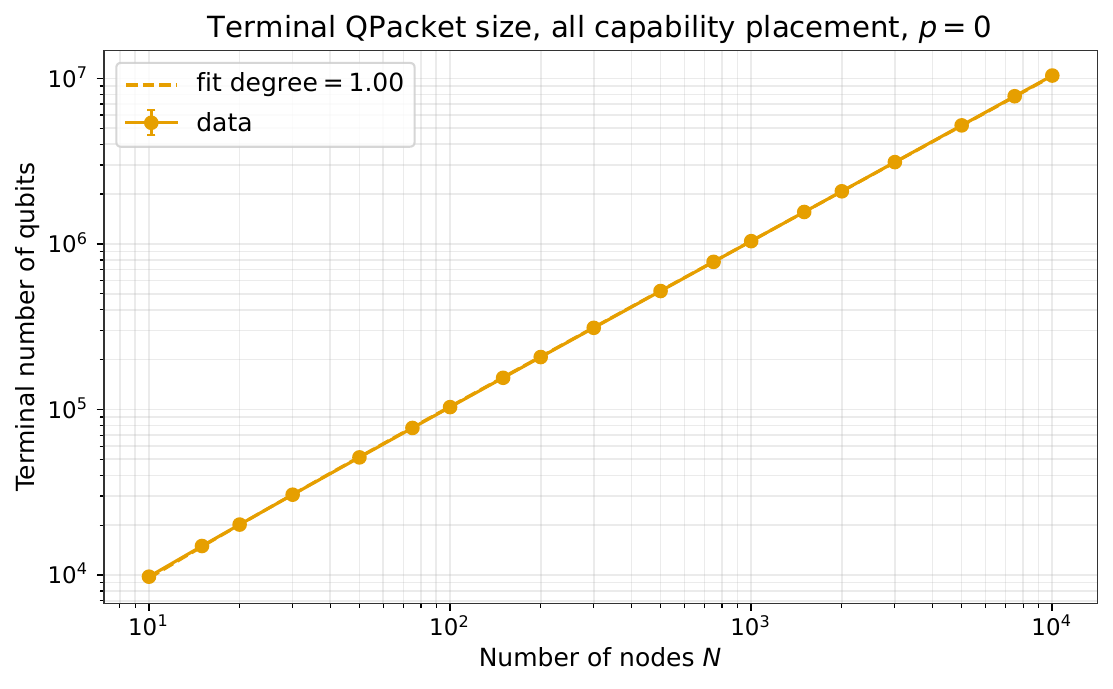}
    \caption{}
    \label{fig:qpacket-terminal-size}
  \end{subfigure}
  \hfill
  \begin{subfigure}[t]{0.33\textwidth}
    \centering
    \includegraphics[width=\linewidth]{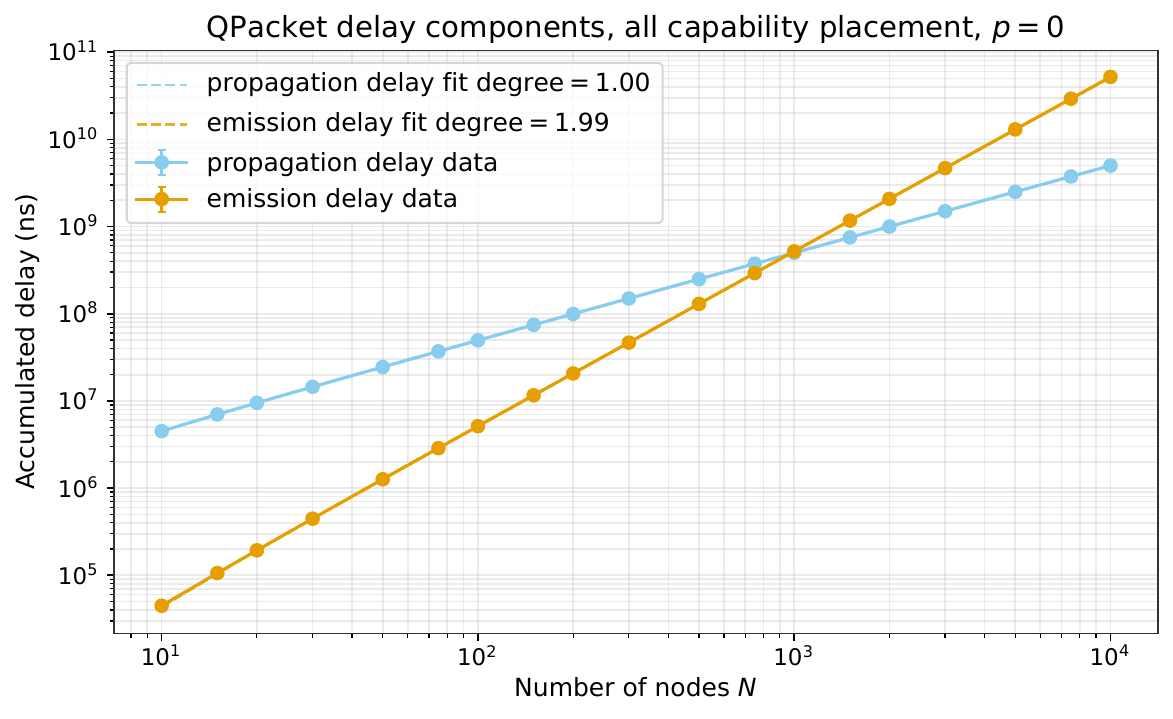}
    \caption{}
    \label{fig:qpacket-delay-components}
  \end{subfigure}
  \hfill
  \begin{subfigure}[t]{0.33\textwidth}
    \centering
    \includegraphics[width=\linewidth]{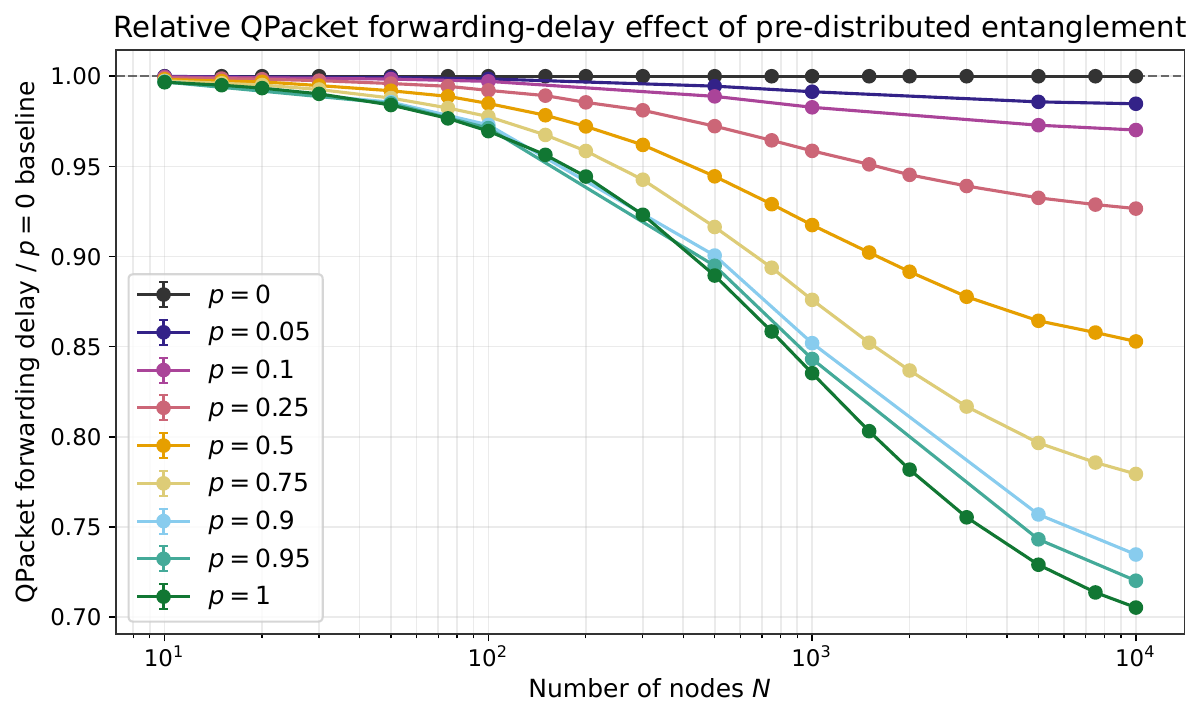}
    \caption{}
    \label{fig:qpacket-relative-delay}
  \end{subfigure}

  \vspace{-0.5em}
  \caption{
  QPacket growth and forwarding cost.
  (a) With \texttt{all} placement and $p=0$, the terminal QPacket size grows approximately linearly.
  (b) Forwarding delay separates into a linear propagation term and a quadratic emission-delay term.
  (c) Increasing pre-distributed entanglement probability $p$ reduces relative forwarding delay by reducing \textsc{LINK\_PREP} stamps.
  }
  \Description{
    Three line plots showing QPacket scaling with network size. The first shows approximately linear growth of terminal QPacket size. The second separates accumulated forwarding delay into a linear propagation component and an approximately quadratic emission component. The third shows that increasing pre-distributed entanglement reduces forwarding delay relative to the case with no pre-distributed entanglement.
    }
  \label{fig:qpacket-scaling}
\end{figure*}
\subsection{Effect of Pre-distributed Entanglement}
\label{sec:3.2}
Here, we evaluate the executable specialization under varying availability of pre-distributed entanglement resources. 

Fig.~\ref{fig:deterministic-predistribute} compares the measured success probability with the theoretical model $p^{n_{\mathrm{gen\text{-}unres}}}$ for deterministic capability placements. The simulations show that failures depend on the link-resolvability condition: any generator-unresolvable physical link must be rescued by pre-distributed entanglement. For \texttt{finalBottleneck}, the success probability follows $p$, since the placement leaves exactly one unresolvable 
link: every node except the last two exposes $\textbf{MP}_{\textbf{GEN}}$.

Fig.~\ref{fig:comm-load-predistributed} shows that the classical communication load depends not only on pre-distributed entanglement, but also on the capability placement, since the placement determines how far the request progresses before termination. For placements that are already link-resolvable at $p=0$, namely \texttt{all} and \texttt{every2}, increasing $p$ mainly replaces runtime link preparation with pre-existing FTQ resources. Consequently, fewer EPR-pair halves are transmitted during the service fulfillment and fewer classical synchronization messages are exchanged. The number of QPacket transmissions remains essentially fixed by the chain length, since successful requests still forward the QPacket across the repeater chain.

For unresolvable-link placements, the interpretation is more subtle. As $p$ increases, pre-distributed entanglement may allow the service intent to survive earlier unresolvable links and fail later, or eventually succeed. The communication load can therefore increase, since more of
the service workflow is actually executed. In this regime, pre-distributed entanglement reduces the need for local link preparation, but it also increases the amount of signaling and forwarding performed before termination for runs that ultimately fail. This is particularly salient from the \texttt{finalBottleneck} scheme in Fig.~\ref{fig:comm-load-predistributed}, which uses nearly the same amount of classical messaging as the successful schemes despite failing. 
This observation again highlights the tight coupling between classical
and quantum dynamics in quantum networks, as well as the impact of
architectural choices. Increasing entanglement-resource availability improves success probability, but the resulting classical signaling load depends on the capability placement, since it determines how far requests progress before succeeding or failing. Thus, the relevant question is not simply whether pre-distributed entanglement increases or reduces classical cost; rather, the cost depends on the interaction among resource availability, node capabilities, and the Dynamic-Kernel pipeline. The instrumentation therefore enables evaluation beyond success/failure outcomes, including classical signaling load, which is natively supported by the ns-3 environment and is essential for realistic quantum-network design.

Lastly, we examine how random generator placement behaves by increasing $p$. Fig.~\ref{fig:random-heatmap} reports
the success probability as a function of $g$ and $p$ for $N = 100$. As expected, both parameters improve success: increasing $g$ reduces the occurrence of
adjacent non-generator nodes, while increasing $p$ covers otherwise unresolvable links with pre-distributed entanglement. The heatmap is therefore governed by the same link-resolvability condition as the deterministic placements, but now with two coverage mechanisms. In the sampled region, the iso-success contours in Fig.~\ref{fig:random-heatmap} vary more sharply along the $g$ axis than
along the $p$ axis. This is expected in the considered worst-case autonomous regime: making one node generator-capable can resolve up to two adjacent physical links, whereas pre-distributed entanglement resolves only the individual link on which it is present. 

\subsection{QPacket Growth}
\label{sec:3.3}
We now focus on the QPacket meta-header growth behavior and its forwarding cost. Figs.~\ref{fig:qpacket-scaling}a--b show the scaling of the QPacket meta-header size and of the accumulated time spent forwarding it. We use the \texttt{all} capability placement with $p=0$, although the same results apply to all successful $p=0$ placements in this case study, since the number of stamps does not change. The scaling follows directly from the committed stamp sequence: Alice appends \textsc{LINK\_PREP} and \textsc{ACT\_FORWARD}, while each intermediate repeater appends \textsc{LINK\_PREP}, \textsc{SWAP}, and \textsc{ACT\_FORWARD}. Hence, the $f \geq 0$-th forwarded QPacket carries $s_f = 2+3f$ stamps. At the final forwarding step to Bob, $f = N - 2$, the terminal number of accumulated stamps is linear in $N$:
\[
s_{N - 2} = 2+3(N-2) = O(N).
\]
Because each forwarding event carries the meta-header accumulated so far, linear terminal growth induces a quadratic accumulated forwarding cost. For an overhead of $c$ qubit-equivalents per stamp, the total QPacket forwarding time is:
\[
\begin{aligned}
&T_{\mathrm{QPacket}}(N) = \sum_{f=0}^{N-2}(t_{\text{delay}} + (cs_f - 1)t_{\text{emit}}) =\\
&(N-1)(t_{\text{delay}} - t_{\text{emit}}) + \left(\frac{3}{2}(N-1)^2+\frac{1}{2}(N-1)\right)c t_{\mathrm{emit}} = O(N^2).
\end{aligned}
\]
To verify this scaling, we fit the simulated accumulated QPacket forwarding time as:
\[
T_{\mathrm{QPacket}}(N)=a (N-1) + b (N-1)^2.
\]
where the linear coefficient captures the per-hop propagation component, and the quadratic coefficient captures the cost of repeatedly forwarding a growing meta-header.

%
The results highlight that the accumulated QPacket forwarding cost is not determined by network size alone. Indeed, increased network resources, such as pre-distributed entanglement and associated policies, can affect the required number of actions and therefore the size and forwarding time of QPackets. 

Fig.~\ref{fig:qpacket-scaling}c shows the accumulated forwarding time relative to the $p = 0$ baseline for the \texttt{all} capability placement, chosen so that failure modes do not factor into this comparison. When $p=1$, adjacent links are already present in FTQ: Alice appends only \textsc{ACT\_FORWARD}, while each repeater appends \textsc{SWAP} and\\
\textsc{ACT\_FORWARD}. The $f$-th forward therefore carries fewer stamps than in the $p=0$ case, i.e., $1+2f$ stamps, giving a quadratic leading term of $(N-1)^2 c t_{\mathrm{emit}}$, which is $2/3$ the size of the leading term for $p=0$. The relative delay in Fig.~\ref{fig:qpacket-scaling}c therefore approaches $2/3$ as $p$ increases, with a slightly larger value due to the additional linear term. Intermediate $p$ values interpolate between these regimes by reducing the expected number of runtime \textsc{LINK\_PREP} stamps. 

While these results are specific to this service request and policy, there is a more general takeaway: QPacket growth is not exclusively a matter of encoding and network size. It also depends on policy choices, resource availability, and the granularity of committed actions. Consequently, QPacket design, policy design, and network
architecture must be evaluated jointly.

\section{Discussion}

This work demonstrated a Q2NS/ns-3 executable specialization of the Dynamic-Kernel/QPacket logic introduced by the beyond-layering protocol suite~\cite{CacCal-26}. The implementation instantiates QPackets carrying the service intent and append-only action-commit stamps, together with a node-local Dynamic Kernel Planner--Executor--Engine processing pipeline. The implementation deliberately targets a scoped service, namely e2e entanglement distribution over a repeater chain, so that the resulting behavior can be instrumented, analytically verifiable, and interpreted in terms of explicit policy choices.

The simulated results match the analysis that for the considered worst-case autonomous regime service-fulfillment success depends on a link-resolvability condition: each link must have pre-distributed entanglement or at least one endpoint capable of entanglement generation. 
This agreement is significant because the implementation includes nontrivial Dynamic-Kernel behavior: MP-agnostic planning, capability-dependent Executor binding, failure stamps for non-bindable \textsc{LINK\_PREP} actions, QPacket forwarding, and local soft-state release.

Instrumentation also clarifies how node capabilities and resource availability shape execution cost, not only success probability. For placements that already succeed without pre-distributed entanglement, increasing $p$ reduces runtime \textsc{LINK\_PREP} and therefore lowers both quantum transmissions and classical synchronization. For placements
with generator-unresolvable links, the effect is placement-dependent: additional pre-distributed entanglement can let a request progress farther before succeeding or failing, thereby exposing later-stage
signaling, forwarding, and soft-state-trigger costs. Thus, the cost of fulfilling a given service intent is not determined by topology, network size, or resource availability in isolation. It emerges from their interaction with node capabilities and node-internal state.
This is exactly the type of nontrivial dependency that the Dynamic-Kernel pipeline captures without global synchronization: each node recomputes its continuation from the QPacket meta-header and its own local state.

The presented executable specialization enables identification of next technical steps such as richer service intents, additional MP-library heterogeneity, and specific QPacket encoding choices. Overall, the paper provides evidence that Q2NS/ns-3 can support reproducible evaluation of quantum-native protocol-suite concepts, not only isolated quantum-network primitives.

\section{Artifact Availability}
\label{sec:artifact}


Q2NS is released through the ns-3 App Store~\cite{q2ns-appstore}. The separate artifact for this paper uses Q2NS as the ns-3-based quantum-network simulation substrate, while providing the Dynamic-Kernel/QPacket specialization. The artifact includes the added C++ implementation logic, data collection scripts, raw CSV outputs, and analysis scripts used to regenerate the main figures, along with a README with step-by-step instructions. The repository also provides a Docker-based workflow, supporting easily reproducible execution of the main experiments. The artifact is available on Zenodo~\cite{Q2NSBeyondLayeringArtifact-26} and GitHub at \url{https://github.com/QuantumInternet-it/q2ns-beyond-layering} under release tag \texttt{icns3-2026-submission}.

\begin{acks}
This work has been funded by the European Union under Horizon Europe ERC-CoG grant QNattyNet, n.101169850. Views and opinions expressed are however those of the author(s) only and do not necessarily reflect those of the European Union or the European Research Council Executive Agency. Neither the European Union nor the granting authority can be held responsible for them.
\end{acks}

\bibliographystyle{ACM-Reference-Format}
\bibliography{references}

\end{document}